\newcommand{\PaperTitle}{\tool: An Automated and Reproducible Environment Toolkit for DNS Protocol Analysis}

\documentclass[10pt,sigconf,letterpaper,nonacm]{acmart}
\usepackage{xspace}

\usepackage{amssymb}  
\usepackage{pifont}  
\usepackage{tabularx}
\usepackage{threeparttable}
\usepackage{graphicx}
\usepackage{subcaption}

\newcommand\tool{{\sc Dart\xspace}}

\begin{document}

\settopmatter{printacmref=false} 
\setcopyright{none}             
\renewcommand\footnotetextcopyrightpermission[1]{} 
\pagestyle{plain}

\title{\PaperTitle}

\author{Yunyi Zhang}
\email{yunyizhang@mail.tsinghua.edu.cn}
\affiliation{%
  \institution{Tsinghua University}
  \city{Beijing}
  \country{China}
}

\author{Xikai Xiong}
\email{xchi_xikai@outlook.com}
\affiliation{%
  \institution{Tsinghua University}
  \city{Beijing}
  \country{China}
  }

\author{Baojun Liu}
\email{lbj@tsinghua.edu.cn}
\affiliation{%
  \institution{Tsinghua University}
  \city{Beijing}
  \country{China}
  }

\author{Haixin Duan}
\email{duanhx@tsinghua.edu.cn}
\affiliation{%
  \institution{Tsinghua University}
  \city{Beijing}
  \country{China}
  }

\begin{abstract}


Domain Name System (DNS) protocol analysis is fundamental to understanding and fortifying the Internet’s naming infrastructure. However, the lack of automated, portable, and user-friendly environment orchestration tools imposes significant overhead on researchers and severely limits the reproducibility of DNS studies. 
In this paper, we present \tool, an automated toolkit specifically engineered for DNS protocol analysis. \tool \  employs a declarative syntax to abstract the intricate software dependencies and heterogeneous configuration requirements of diverse DNS implementations, providing a unified and streamlined orchestration interface.
We describe the architecture of \tool \ and evaluate its performance.
Furthermore, we present two case studies that illustrate how \tool \ enables researchers to construct portable DNS analysis environments with a single command. 
To foster transparency and facilitate future research, all replicated environments and configurations will be open-sourced.

\end{abstract}

\maketitle

\section{Introduction}

The Domain Name System (DNS) is a critical component of the Internet, supporting numerous network services, like Email, Web, and CDN~\cite{Amazon-CloudFront}. 
As evidenced by several large-scale Internet outages and high-profile service hijackings, DNS remains notoriously complex to manage. Systematic DNS protocol analysis is therefore essential for the community to characterize the evolving DNS ecosystem and mitigate the impact of potential risks.

A fundamental prerequisite for DNS protocol analysis, especially for new risks discovery, is the construction of robust testing and validation environments. Such efforts often require large-scale evaluation across multiple DNS implementations to assess the impact scope of a vulnerability. For instance, TsuKing~\cite{TsuKing2023} needs to build a network topology with dozens of nodes based on four different implementations. These environments are almost exclusively built manually by researchers and are typically not open-sourced. As shown in Section~\ref{sec:rw}, our analysis of DNS-related papers published at top-tier security venues in recent years reveals that only 20\% (4/20) have open-sourced their environments. 
Such limitations are rarely intentional. Rather, researchers are often overwhelmed by the heterogeneous configuration requirements and complex dependencies of diverse software. Moreover, these manually crafted environments frequently lack the portability necessary for open-source distribution, thus hindering the realization of reproducible research.

In this paper, we introduce \tool, an automated and reproducible environment toolkit for DNS protocol analysis. 
\tool \ is composed of: (1) \textit{DNS FS}, a specialized resource management layer hosting an extensive library of 426 versions across four mainstream resolvers (e.g., BIND~\cite{bind}, PowerDNS~\cite{powerdns}) to ensure environment consistency; (2) \textit{DNS Builder}, the core orchestration engine that implements the Configuration-as-Code (CaC) paradigm to abstract heterogeneous software dependencies and configurations into a unified, declarative syntax; (3) \textit{DNS SAK}, a versatile module dedicated to generating diverse protocol testing payloads; and (4) \textit{DNS Monitor}, an observability module that leverages eBPF-based instrumentation to provide high-fidelity, real-time insights into cache, resolution path, CPU utilization, and processing latency. Crucially, \tool \ achieves automated configuration for DNSSEC, a pivotal advancement that significantly lowers the barrier to entry for DNSSEC-related research.

We evaluated \tool \ to demonstrate its efficacy in significantly alleviating the orchestration challenges faced by researchers. First, we utilized \tool \ to replicate 20 representative DNS research environments. Most of these setups were achieved with approximately 100 lines of configuration; notably, the complex TsuKing environment, comprising dozens of resolvers, was orchestrated using only 66 lines, reducing redundant effort by over 90\% in configuration overhead compared to native Docker Compose. Second, our results confirm that the observability module introduces negligible performance overhead to the resolvers. Furthermore, \tool's enhanced packet manipulation capabilities, such as IP fragmentation and stateful protocol interaction, provide researchers with advanced primitives for DNS analysis.

Moreover, we complement our evaluation with two case studies that explore internal resolution behaviors across software versions and assess RFC compliance across mainstream open-source resolvers and proprietary commercial implementations. Our findings emphasize that \tool \ serves as a powerful enabler for large-scale, cross-implementation analysis, revealing behavioral disparities that were previously difficult to quantify. 
By providing a unified testing framework, \tool \ empowers researchers to systematically investigate the vast and heterogeneous DNS resolver landscape with unprecedented ease and precision.

\noindent \textbf{Contributions.} Our contributions are outlined below:

\textit{Toolkit.} We develop the first automated orchestration toolkit for DNS protocol analysis. It abstracts implementation heterogeneities into a unified, declarative framework, significantly reducing the configuration overhead for researchers.

\textit{Reproducible Testbeds.} Using \tool, we replicate the experimental environments of 20 DNS studies (e.g., cache poisoning, DoS). These standardized environments will be open-sourced to provide a baseline for reproducible DNS research~\footnote{https://anonymous.4open.science/r/DART-1D8C}.
\section{Related Work}
\label{sec:rw}

\noindent \textbf{DNS protocol analysis.}
Current DNS research primarily focuses on discovering vulnerabilities such as cache poisoning~\cite{Kaminsky2010,ForwardFrag2020,Inject2021,Maginot2023,PhoenixDomain2023,TuDoor2024,DgradDNSSEC2023,Rebirthday2025} and DoS attacks~\cite{NXNS2020,TsuName2021,TsuKing2023,NRDelegation2023,TuDoor2024,Loopy2024,KeyTrap2024,CAMP2024,DNSBomb2024}, where researchers predominantly rely on the manual construction of test environments across various software implementations to validate attack efficacy.
However, even recent automated analysis attempts like ResolverFuzz~\cite{ResolverFuzz2024}, remain constrained by rigid experimental environments. 
Our review of typical DNS studies from the past five years reveals that $16 / 20$ works fail to provide a functional or reproducible experimental environment, highlighting the urgent need for standardized testing frameworks.

\noindent \textbf{Network testbeds.}
The demand for reproducible research has driven the development of various network testbeds and automated toolkits. Existing network testbeds range from general-purpose emulators like Mininet~\cite{Mininet2010}, GNS3~\cite{gns3}, and EVE-NG~\cite{EVE-NG} to infrastructure-focused environments such as SEED~\cite{SEED2022}, which primarily serve topological validation and pedagogical purposes. Recent advancements have introduced domain-specific platforms like the Gotham~\cite{CamaraFAUZ24} for IoT research, as well as cloud-native ``Lab-as-Code'' solutions like Containerlab~\cite{containerlab} for rapid container orchestration. 
However, these platforms remain largely protocol-agnostic; they prioritize general-purpose, still requiring DNS researchers to expend significant manual effort in constructing specialized experimental environments.



\section{\tool: DNS Analysis and Reproduction Toolkit}
In this section, we describe \tool's requirements and principles. Then, we introduce the architecture.

\subsection{Requirements and Design Principles}
The goal of \tool \ is to transform the environment setup in DNS protocol analysis from a manual, ad hoc process into a portable and automated workflow.
Based on the challenges in DNS protocol analysis and real-world DNS software deployment, we summarize the core requirements for DART.

\textbf{Usability.} To facilitate large-scale testing, the framework must provide environment transparency. Researchers should be able to rapidly configure target resolver environments without managing low-level network topologies, divergent configuration syntaxes across DNS software suites, or the intricate dependency chains associated with software deployment, particularly for legacy versions.

\textbf{Extensibility}. To accommodate specific threat scenarios, the framework must enable researchers to craft customized payloads, such as IP fragmentation and tailored authoritative responses. Furthermore, it should provide fine-grained control and monitoring over DNS software behaviors, such as the seamless modification of root hints.

\textbf{Portability.} To ensure rigor and scientific reproducibility, generated testing environments must be decoupled from host-specific dependencies. The framework should implement a "build once, reproduce anywhere" paradigm, ensuring that experimental results remain consistent across heterogeneous computing infrastructures regardless of local environmental parameters.

\subsection{Architecture}

The design of \tool \ is underpinned by the principle of \textit{Con} \textit{figuration-as-Code.} By adopting a declarative construction paradigm, it transforms high-level user requirements into a comprehensive set of configurable primitives. Furthermore, \tool employs parameterized abstractions to shield researchers from cross-software heterogeneities, thereby automating repetitive operational tasks and the orchestration of complex network topologies.

The \tool \ is composed of: (1) DNS FS, a decoupled repository for managing images and system resources; (2) DNS Builder, the orchestration engine for automated environment provisioning; (3) DNS SAK, an integrated suite of specialized testing utilities; and (4) DNS Monitor, which provides fine-grained state monitoring of target resolvers.

\subsubsection{DNS FS}
DNS FS is an architectural isolation layer positioned between the build engine and the underlying physical environment. Serving as a secure sandbox and staging area for both remote images and local assets, it leverages a custom path model (DNSBPath) to facilitate multi-protocol distribution, cache management, and resource isolation. Finally, DNS FS enabled \tool \ to automate dependency configurations for 426 versions across four mainstream resolvers, ensuring high-fidelity environment consistency.

\subsubsection{DNS Builder}

At the heart of \tool \ lies DNS Builder, a declarative, configuration-as-code driven engine designed for the automated synthesis of DNS environments. By leveraging YAML-based specifications, researchers can define complex network topologies and protocol configurations at a high level of abstraction. The engine then autonomously orchestrates the deployment process, and inject customized configurations to instantiate a ready-to-test environment.

Next, we show the seven key capabilities of DNS Builder that enable seamless and portable DNS testing.

\textbf{Behavior Primitives.}
%
To shield researchers from configuration disparities and enable them to focus on experimental design, \tool \ introduces a behavioral modeling layer that abstracts DNS operations into high-level behavioral primitives (e.g., RootHint, Forward, Stub, and Master). During the construction phase, the engine dynamically maps these primitives to the specific implementation logic of the target DNS software, automatically provisioning the necessary configuration files and resource dependencies. 

\textbf{Reference Pointer.}
To mitigate redundant operations when orchestrating large-scale environments with multi-tier resolvers, \tool \ introduces \textit{Reference Pointer}.
Specifically, the framework supports a modular configuration structure where a master profile can import external configurations via include directives. Furthermore, images and DNS services within a profile can inherit attributes from built-in templates or sibling configurations using the \texttt{ref} pointer. 
The DNS Builder engine then autonomously executes a three-stage resolution process: (1) importing external references, (2) flattening hierarchical structures into a unified representation, and (3) resolving conflicts during attribute merging. 

\textbf{Configuration Hot Loading.}
DNS Builder implements a suffix-based automated discovery mechanism to support incremental configuration hot loading. It autonomously identifies supplemental DNS configuration snippets based on predefined file suffixes and dynamically injects the corresponding references into the main configuration.

\textbf{Placeholder.}
To enhance configuration flexibility, \tool \ provides a rich set of built-in placeholders, such as \texttt{proj.inet} for project subnets and \texttt{builds.root.ip} for root server addresses. 
This mechanism decouples service configurations from static environment parameters, significantly bolstering the portability and robustness of the testing environment.

\textbf{Auto Hook.}
To extend the power of static configurations, \tool \ integrates auto hooks at the environment construction process, enabling the dynamic modification of configurations via custom scripts. This mechanism effectively overcomes the inherent limitations of static YAML specifications.

\textbf{Automated DNSSEC Signing and Deployment.}
Traditionally, constructing a functional DNSSEC environment is a laborious process, plagued by the complexities of manual key management and inter-registry coordination. To address this, \tool \ streamlines the entire DNSSEC lifecycle during the construction phase: it automates the generation or reference of DNSKEYs, performs autonomous zone signing, and establishes the chain of trust originating from the root. By abstracting away these intricate cryptographic interdependencies, DART significantly reduces the operational barrier for researchers to evaluate DNSSEC-enabled infrastructures.

\textbf{Pluggable Resources.}
DART offers pluggable resources loading that enables researchers to encapsulate the behaviors of arbitrary DNS implementations, including proprietary, closed-source commercial solutions, using standardized interface classes. Upon initialization, DNS Builder autonomously inspects the environment to discover installed Python packages, dynamically loading validated plugin resources via their \texttt{on\_load} hooks. This design achieves implementation transparency, empowering users to evaluate closed-source DNS software as its open-source counterparts.


\subsubsection{DNS SAK}
To address diverse requirements for customized query and response testing, \tool \ incorporates the DNS SAK (Swiss Army Knife) module. This component provides researchers with a robust suite of testing payloads and capabilities, including: (i) programmable query and response generation for both clients and authoritative servers; and (ii) support for advanced attack primitives, such as IP fragmentation, IP spoofing, and high-speed packet injection.

\subsubsection{DNS Monitor}

Effective DNS protocol analysis requires insights into why a resolution failed, rather than merely identifying its outcome. While traditional tools like dig treat resolvers as black boxes, capturing only high-level RCODEs (e.g., SERVFAIL), \tool \  leverages eBPF-based instrumentation to monitor four critical metrics: cache states, resolution logic, CPU load, and processing latency. Furthermore, the framework supports extensible observation modules, allowing researchers to implement custom probes for specialized monitoring requirements.

\begin{table}[]
\caption{Summary of DNS risks replicated via \tool.}
\label{tab:threats}
\scalebox{0.75}{
\begin{threeparttable}
\begin{tabular}{@{}cccccc@{}}
\toprule
\textbf{}     & \textbf{Year} & \textbf{Software} & \textbf{Version} & \textbf{LoC} & \textbf{\#Resolver $^*$} \\ \midrule
Birthday~\cite{Birthday}      & 2002             & BIND                 & 9.4.3                & 42              & 3                   \\
Kaminsky~\cite{Kaminsky2010}      & 2008          & BIND              & 9.4.3            & 38           & 3                   \\
Forward Frag.~\cite{ForwardFrag2020} & 2020          & Dnsmasq           & 2.82             & 78           & 5                   \\
NXNSAttack~\cite{NXNS2020}          & 2020          & Unbound           & 1.10.0           & 107          & 4                   \\
TsuName~\cite{TsuName2021}       & 2021          & BIND                 & 9.18.0                & 81           & 6                   \\
Inject~\cite{Inject2021}        & 2021          & Unbound                 & 1.19.0                & 73           & 5                   \\
Maginot~\cite{Maginot2023}       & 2023          & BIND              & 9.18.0           & 159          & 6                   \\
TsuKing~\cite{TsuKing2023}       & 2023          & Unbound           & 1.17.1           & 66           & 2*n + 2             \\
NRDelegation~\cite{NRDelegation2023}  & 2023          & Unbound           & 1.16.0           & 111          & 5                   \\
PhoenixDomain~\cite{PhoenixDomain2023} & 2023          & Unbound           & 1.16.1           & 125          & 4                   \\
TuDoor~\cite{TuDoor2024}        & 2023          & BIND              & 9.18.14          & 84           & 3+n                 \\
Dgrad. DNSSEC~\cite{TuDoor2024} & 2023          & BIND              & 9.11.3           & 52           & 4                   \\
Loopy~\cite{Loopy2024}         & 2024          & hickory~\cite{hickory}  & 0.22.0                & 39           & 2                   \\
KeyTrap~\cite{KeyTrap2024}       & 2024          & Unbound           & 1.19.0           & 93           & 4                   \\
CAMP~\cite{CAMP2024}          & 2024          & BIND               & 9.18.4               & 271          & f(n)                \\
DNSBomb~\cite{DNSBomb2024}       & 2024          & Unbound           & 1.19.0           & 189          & 4                   \\
Rebirthday~\cite{Rebirthday2025}    & 2025          & Unbound           & 1.22.0           & 122          & 4                   \\
DNSPUN~\cite{DNSPUN}   & 2026          & Unbound           & 1.22.0           & 116          & 4                   \\
BADDNS~\cite{BADDNS}    & 2026          & PowerDNS          & 5.2.2            & 70           & 5                   \\
Cuckoo~\cite{Cuckoo}       & 2026          & BIND                 & 9.20.3                & 108          & 4                   \\ \bottomrule
\end{tabular}

\begin{tablenotes}
        \footnotesize
        \item  *: The $n$ represents the resolver scale, allowing the number of resolvers to be increased simply by modifying a single configuration value. 
    \end{tablenotes}
\end{threeparttable}

}
\end{table}


\section{Evaluation}

In this section, we evaluate the efficiency, reliability, and observational overhead of \tool. To demonstrate its flexibility in orchestrating complex testing environments, we leverage DART to reproduce 20 representative DNS attacks, spanning cache poisoning and DoS. Furthermore, we provide a comparative analysis between DART and traditional environment construction frameworks. 


\subsection{Efficiency and Reliability}
\label{subsec:er}
\noindent \textbf{Efficiency evaluation.} 
The core objective of \tool \ is to help researchers rapidly build reproducible environments for DNS protocol analysis. 
To evaluate this, we first test \tool’s orchestration capabilities in various analysis scenarios, focusing on the following three aspects: (i) Handling system dependencies for legacy resolvers. 
(ii) Scalability in complex network topologies. 
(iii) Automated DNSSEC provisioning. 

Specifically, we selected 20 representative DNS vulnerabilities, covering cache poisoning and DoS attacks. Table~\ref{tab:threats} summarizes our reproduction results. Using \tool, we can easily construct diverse DNS testing environments. 
For Kaminsky, \tool \ helps researchers bypass complex underlying system dependency issues through high-level abstractions and pre-configured images. 
Moreover, the results show that \tool \ requires only a 66-line declarative file to build a network topology containing 34 resolvers.
Finally, \tool \ enables the construction of a top-down chain of trust within its closed DNS environment. Researchers can replicate the environment for KeyTrap~\cite{KeyTrap2024} using only 93-line configuration.

More importantly, these environments are highly portable; once \tool \ is deployed, the target environment can be constructed with a single command. To further support the reproducibility of DNS protocol analysis, we will open-source the environments we have implemented.

\noindent \textbf{Reliability evaluation.}
To avoid ethical concerns and uncontrollable Internet noise, DNS protocol analysis must be conducted in isolation. We therefore assess \tool's ability to provide a reliable and fully contained DNS environment.


Specifically, we selected the top 1,000 domains in Tranco~\cite{Tranco2023} as our test domains. We then used \tool \ to construct a test environment, comprising a recursive resolver (BIND 9.18.0) alongside virtual Root, TLD, and SLD nameservers (BIND 9.20.3). Leveraging the Auto Hook and Behavior Primitives, we batch-injected these domains into the environment with unified response policies. Finally, we performed bulk resolution of these domains at the client side to observe whether any resolution traffic leaked into the Internet.

The test results show that all tested domains received the pre-configured responses, with the entire resolution process successfully contained within the isolated DNS environment. This demonstrates that the independent DNS constructed by \tool \ provides high reliability and absolute isolation. 

\begin{figure}[h]
    \centering
    \includegraphics[width=1.0\linewidth, trim=0cm 0.3cm 0cm 0.2cm, clip]{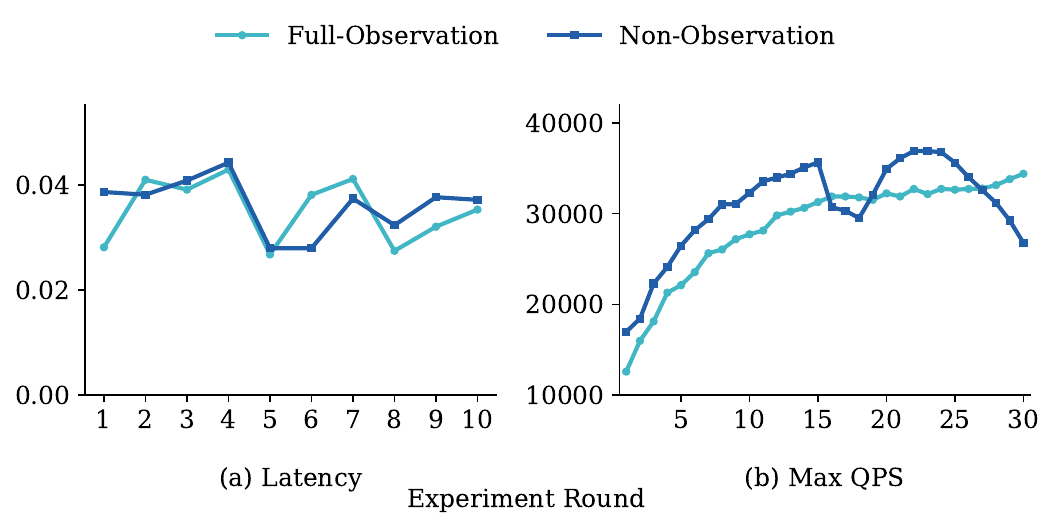}
    \caption{Observational overhead evaluation results.}
    \label{fig:overhead}
\end{figure}

\subsection{Observational Overhead}
In this section, we evaluate the actual performance overhead imposed by the observation module on the target resolver.

Under identical hardware conditions (32 virtual cores \& 32GB RAM), 
We evaluated \tool \ across two configurations: Non-Observation (module disabled) and Full Observation (module enabled). A BIND 9.20.3 recursive resolver with default settings was deployed in both. Using Tranco Top-1M domains as the query workload, we conducted ten trials per experiment to mitigate stochastic noise such as network latency.
Then, we first measured the observation module's impact on latency by querying resolvers at a sustained 500 QPS for ten 60-second rounds. We then evaluated peak throughput by incrementally increasing the request rate from zero until the resolver's success rate first dropped below 100\%, identifying the maximum sustainable QPS.

\noindent \textbf{Results.} 
The observation module introduces negligible latency overhead and maintains throughput within 7\% of the baseline. As shown in Figure~\ref{fig:overhead}(a), the average latency in Full Observation Mode (0.0351s, $\sigma$=0.08) was statistically comparable to Non-Observation Mode (0.0361s, $\sigma$=0.08). The marginal improvement in observation mode is attributed to inherent network jitter rather than systematic bias, confirming that \tool\ imposes no perceptible delay on DNS resolution.
Figure~\ref{fig:overhead}(b) shows the peak throughput distribution, where the system reached 30,698 QPS in Non-Observation Mode versus 28,607 QPS in Full Observation Mode. This marginal 6.8\% decrease confirms the observation mechanism's efficiency, ensuring it remains a non-bottleneck for large-scale parallel testing.


\subsection{Granularity of Operation}
\tool \ provides researchers with flexible control over DNS packets. We compared the operational granularity of \tool \ against classic tools such as zdns, dig, Scapy, and hping3 in constructing test payloads.

As shown in Table~\ref{tab:tool}, existing solutions present a trade-off: high-concurrency measurement tools (e.g., ZDNS, dnsperf) lack full-stack, cross-layer control, while flexible libraries like Scapy suffer from low execution efficiency. \tool \ bridges this gap by providing a specialized framework for DNS protocol analysis that harmonizes high concurrency, fine-grained cross-layer control, and stateful interaction.

\begin{table}[]
\caption{Capability Comparison Between DART and Other DNS Tools.}
\label{tab:tool}
\scalebox{0.65}{
\begin{threeparttable}
\begin{tabular}{@{}cccccc@{}}
\toprule
\textbf{Tool} & \textbf{\begin{tabular}[c]{@{}c@{}}Standard\\ Query\end{tabular}} & \textbf{\begin{tabular}[c]{@{}c@{}}Batch \\ Sending\end{tabular}} & \textbf{\begin{tabular}[c]{@{}c@{}}IP\\ Fragment\end{tabular}} & \textbf{\begin{tabular}[c]{@{}c@{}}Flag\\ Manipulation\end{tabular}} & \textbf{\begin{tabular}[c]{@{}c@{}}Stateful\\ Interaction\end{tabular}} \\ \midrule
dig~\cite{dig}           & \ding{52}                      & \ding{56}                      & \ding{56}                    & $\triangle$                & \ding{56}                    \\
dnsperf~\cite{dnsperf}       & \ding{52}                      & \ding{52}                     & \ding{56}                    & \ding{56}                 & \ding{56}                    \\
Scapy~\cite{scapy}         & \ding{52}                      & $\triangle$                     & \ding{52}                   & \ding{52}                & $$\ding{52}$$                    \\
PacketSender~\cite{packetsender}  & \ding{52}                      & $\triangle$                     & \ding{56}                    & \ding{52}                & \ding{56}                    \\
hping3~\cite{hping3}        & $\triangle$                      & \ding{52}                     & \ding{52}                   & \ding{52}                & \ding{56}                    \\
Flamethrower~\cite{flamethrower}  & \ding{52}                      & \ding{52}                     & \ding{56}                    & \ding{56}                 & \ding{56}                    \\
ZDNS~\cite{ZDNS2022}          & \ding{52}                      & \ding{52}                     & \ding{56}                    & \ding{56}                 & \ding{56}                    \\ 
\tool          & \ding{52}                     & \ding{52}                     & \ding{52}                   & \ding{52}                & \ding{52}          \\
\bottomrule
\end{tabular}
\begin{tablenotes}
        \footnotesize
        \item  \ding{56}: Support. $\ding{52}$: Not support.  $\triangle$: Limited support.
    \end{tablenotes}
\end{threeparttable}
}
\vspace{-2mm}
\end{table}

\subsection{Compared with Other Frameworks}
Although dedicated environments specifically built for DNS protocol analysis and research are rare, \tool \ is not the first system designed to construct experimental network environments. Building upon the replication of typical DNS risks discussed in Section~\ref{subsec:er}, we compare \tool's environment construction capabilities with mainstream tools such as EVE-NG, SEED, Containerlab, and Docker Compose.

\noindent \textbf{Support for Diverse DNS Software.} 
SEED provides limited support for high-level application environments, particularly for DNS, where it only supports default BIND versions on Linux. 
While EVE-NG, Containerlab, and Docker Compose allow users to import external images, they require manual resolution of system dependencies. In contrast, DART optimizes the construction of DNS testing environments by supporting the automated deployment of over 426 versions across 4 major DNS resolvers. This significantly reduces the overhead of environment configuration for researchers.

\noindent \textbf{Scalability in Complex Network Topology Construction.} 
While EVE-NG’s graphical user interface is intuitive for beginners, constructing large-scale network topologies requires researchers to perform a significant amount of repetitive manual clicking. 
Similarly, although Containerlab and Docker Compose can build environments based on Dockerfiles, they still necessitate the manual creation of configuration files for each individual node. 
In contrast, \tool \ empowers researchers to automate the construction of large-scale, complex topologies through the use of behavioral semantics, reference pointers, and placeholders. 

\noindent \textbf{Native DNSSEC Support.} 
Existing tools such as EVE-NG, Containerlab, and Docker Compose lack native support for DNSSEC, leaving researchers to grapple with complex and error-prone configuration challenges. In contrast, \tool \ provides built-in automated DNSSEC configuration. Through its "auto-hook" mechanism, \tool \ facilitates DNSSEC threat analysis across various sophisticated scenarios, such as the KeyTrap vulnerability, by automating the intricate process of key generation, signing, and chain-of-trust establishment.

\section{Case Study}
This section shows \tool’s capability to empower researchers in analyzing resolver behaviors, facilitated by its agile orchestration and observability.

\subsection{Differential Analysis of Resolution Behaviors Across Software Versions}

We selected 4 mainstream open-source resolvers, including BIND (9.18.0 \& 9.21.15,), Unbound (1.17.1 \& 1.24.2), PowerDNS (4.5.4 \& 5.2.6), and Knot Resolver~\cite{knotresolver} (5.5.2 \& 6.0.16), as our test subjects. For each software, two distinct versions were chosen to be rapidly deployed via \tool. 
Then, selecting the Tranco Top 1K domains as test domains, we leveraged \tool's observation module to collect and analyze the behavioral data of each resolver.

\begin{figure}[t]
\centering
    \begin{subfigure}{.5\textwidth}
        \centering
        \includegraphics[width=0.6\linewidth]{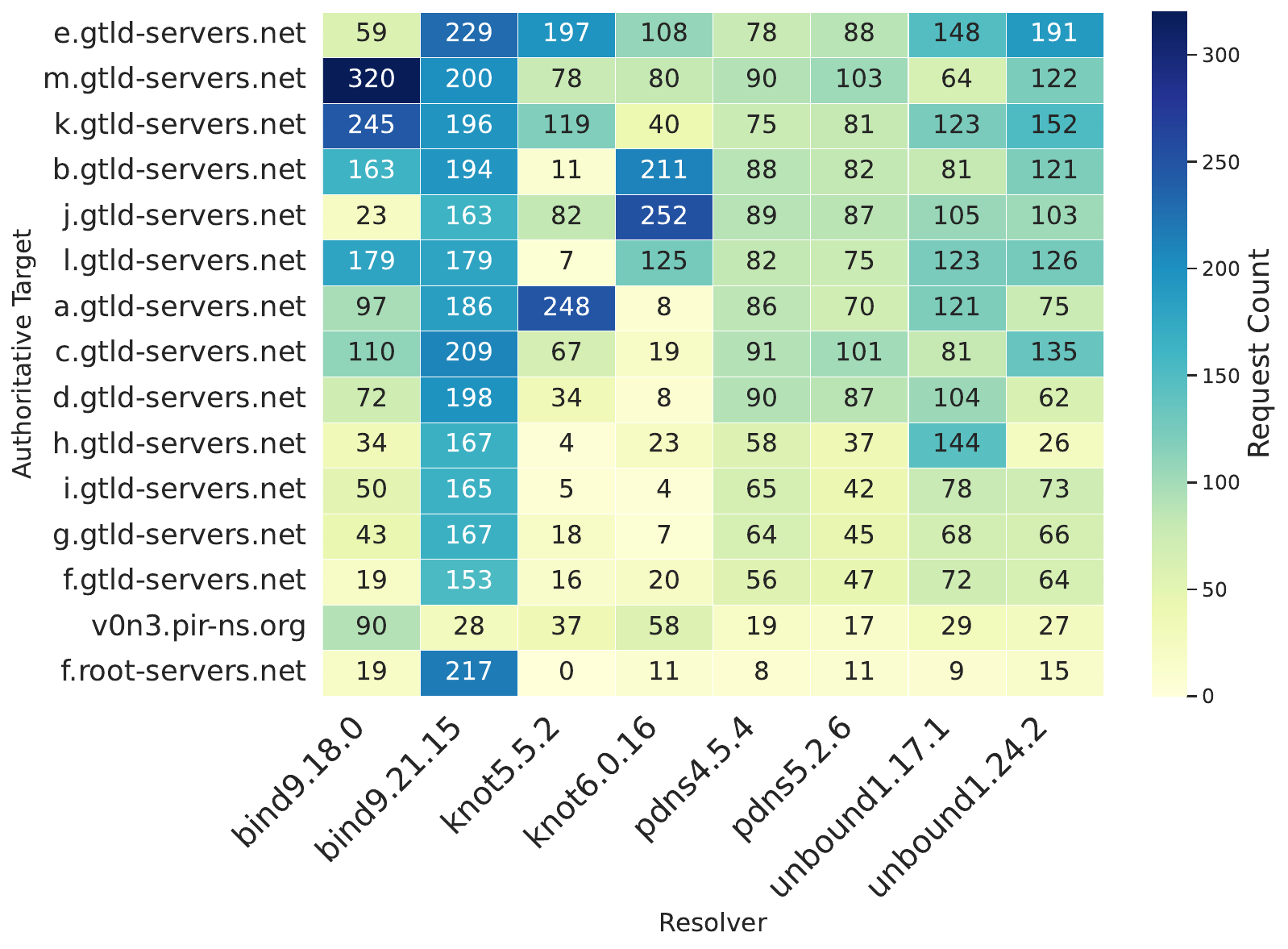}
        \caption{Upstream NS distribution by different implementations.}
        \label{fig:ns_distribution}
    \end{subfigure}\hfill
    \begin{subfigure}{.5\textwidth}
        \centering
        \includegraphics[width=0.7\linewidth]{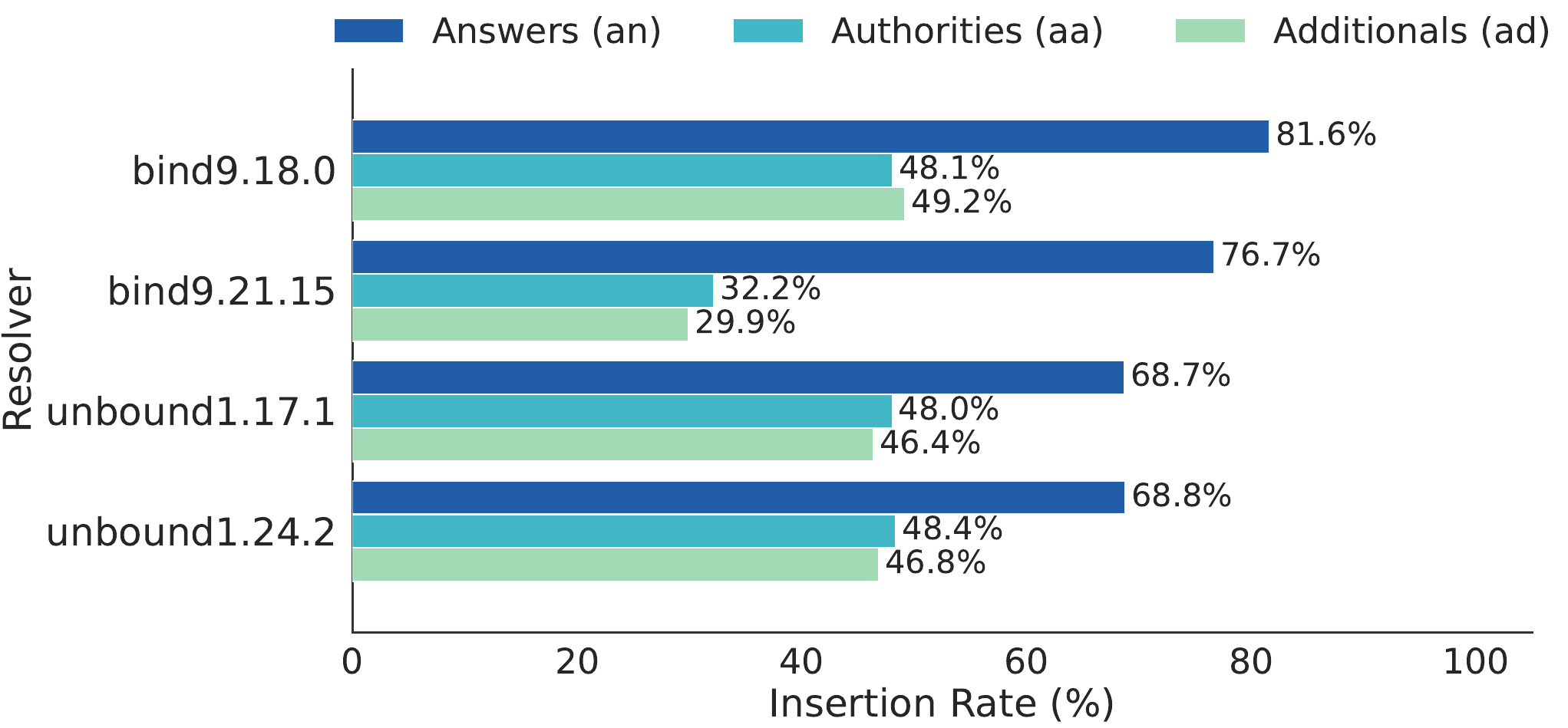}
        \caption{Distribution of cache sources for different implementations.}
        \label{fig:cache_sources}
        \vspace{-2mm}
    \end{subfigure}\hfill

\caption{Analysis of resolution behaviors across software versions.}
\label{fig:resolver_info}
\vspace{-4mm}
\end{figure}

\noindent \textbf{Discrepancy in NS Selection.} 
By comparing resolution workflows under identical configurations, we identified significant disparities in NS load-balancing. As shown in Figure~\ref{fig:ns_distribution}, Knot Resolver diverges from others by directing 95\% of traffic to the fastest NS based on RTT history, with the remaining 5\% randomly distributed. Other resolvers, however, favor a more uniform distribution. We further conducted a source code analysis to validate these findings. Knot utilizes Epsilon-Greedy routing to prioritize low-latency nodes, whereas BIND augments its SRTT algorithm with exponential smoothing, automatic decay, and adaptive quota adjustments to balance the load.

\noindent \textbf{Analysis of Caching Policies.} As shown in Figure~\ref{fig:ns_distribution}, the BIND (v9.21.15) generates significantly higher query volume than its predecessors and other resolvers. Leveraging DART’s cache observation module, we identified the cause: the BIND (v9.21.15) distrusts records in the \texttt{Additional} section of responses, preferring to actively query the IP of each NS. This results in several times more resolution requests under identical configurations. In contrast, most resolvers (e.g., Unbound~\cite{unbound} and BIND (v9.18.0)) directly accept and utilize the glue records provided in the \texttt{Additional} section.


We further analyzed how different resolvers update their caches. Compared to its predecessor, the BIND (v9.21.15) exhibits a significantly lower caching rate for records in the \texttt{Authority} and \texttt{Additional} sections, with a reduction of nearly 20\%, as shown in Figure~\ref{fig:cache_sources}. In contrast, other resolvers like Unbound consistently favor caching and utilizing records from these sections.

Moreover, we investigated cache refresh policies. 
Our findings indicate that most implementations adhere to a partial-order update policy based on record trustworthiness, where existing records are only refreshed by data from the same or a higher trust-level section. For instance, an entry originally cached from the \texttt{Additional} section is updated when the same data appears in the \texttt{Answer} section.



\subsection{Software's RFC Compliance Analysis}
\label{subsec:rfc-compliance}

\tool’s flexible environment orchestration provides an ideal foundation for large-scale security research, particularly for fuzzing frameworks like ResolverFuzz. 
In this section, we evaluate \tool's viability in facilitating differential analysis across DNS implementations. 
Based on our analysis of RFC specifications, we developed ten diverse test cases (T1–T10) covering: functional consistency (T1–T4, e.g., DNAME support, RD flag handling), special-use domain processing (T5–T7, e.g., .test, .onion), and DNSSEC (T8–T10, e.g., anomalous TSIG handling). Detailed case descriptions are provided in Appendix~\ref{ap:test_case}.
Then, we leveraged \tool\ to automate the construction of a comprehensive testbed comprising four open-source DNS resolvers and one commercial implementation (a dedicated service developed by our industry partner using \tool's \textit{Pluggable Resources}).


\begin{table}[]
\caption{Results of differential response analysis.}
\label{tab:cases}
\scalebox{0.7}{
\begin{threeparttable}
\begin{tabular}{@{}ccccccc@{}}
\toprule
\textbf{No.} & \textbf{Name}                        & \textbf{BIND} & \textbf{Unbound} & \textbf{\begin{tabular}[c]{@{}c@{}}PowerDNS\\ Recursor\end{tabular}} & \textbf{\begin{tabular}[c]{@{}c@{}}Knot\\ Resolver\end{tabular}} & \textbf{XDNS} \\ \midrule
T1           & ns\_delegation              & \ding{52}              & \ding{52}                & \ding{52}                          & \ding{52}                      & \ding{52}              \\
T2           & response\_no\_qr & \ding{56}              & \ding{52}                & \ding{52}                          & \ding{56}                      & \ding{56}              \\
T3           & dname\_query                & \ding{56}              & \ding{52}                & \ding{52}                          & \ding{52}                      & \ding{56}              \\
T4           & rd\_flag\_clear             & \ding{56}              & \ding{56}                & \ding{52}                          & \ding{56}                      & \ding{56}              \\
T5           & invalid\_domain         & \ding{56}              & \ding{52}                & \ding{56}                          & \ding{52}                      & \ding{56}              \\
T6           & test\_domain            & \ding{56}              & \ding{52}                & \ding{56}                          & \ding{52}                      & \ding{56}              \\
T7           & onion\_domain               & \ding{56}              & \ding{52}                & \ding{56}                          & \ding{52}                      & \ding{56}              \\
T8           & cd\_flag\_query             & \ding{52}              & \ding{52}                & \ding{52}                          & \ding{52}                      & \ding{52}              \\
T9           & edns\_key\_tag      & \ding{56}              & \ding{56}                & \ding{56}                          & \ding{56}                      & \ding{56}              \\
T10          & multi\_tsig\_records        & \ding{52}              & \ding{56}                & \ding{56}                          & \ding{56}                      & \ding{56}              \\ \bottomrule
\end{tabular}

\begin{tablenotes}
        \footnotesize
        \item BIND: v9.21.15. Unbound: v1.24.2. PowerDNS Recursor: v5.2.6. Knot Resolver: v6.0.16
        \item XDNS is a commercial DNS software from a globally renowned public DNS provider.
    \end{tablenotes}
\end{threeparttable}
}
\vspace{-4mm}
\end{table}

We found that mainstream resolvers still exhibit significant non-compliance with RFC specifications, alongside notable behavioral disparities between different implementations. Table~\ref{tab:cases} summarizes the results of our ten test cases. For instance, in T4, according to RFC standards, a resolver receiving a query without the RD (Recursion Desired) flag should return a REFUSED response accompanied by Extended DNS Error (EDE) Code 20; however, BIND, Knot, and Unbound all opted to proceed with the resolution. While this study presents only a subset of potential scenarios, \tool’s flexible environment orchestration enables researchers to conduct far more extensive and systematic analyses.

For commercial DNS software, our results show that much like its open-source counterparts, it also exhibits significant non-compliance with RFC specifications. For instance, in T2, whereas RFC standards mandate that resolvers must verify the QR (Query/Response) bit in authoritative responses, the provider's implementation failed this check. We have disclosed and discussed these discrepancies with our collaborators, who are currently evaluating these non-compliances.





\section{Conclusion and Future Work}


\tool \ is a declarative and open-source toolkit designed to simplify the orchestration of complex DNS testing environments. By a unified syntax, it enables researchers to deploy diverse, multi-implementation scenarios with minimal configuration, while featuring fully automated DNSSEC provisioning. We hope that \tool's will empower the researchers to more effectively analyze, and secure the DNS ecosystem.

Moving forward, \tool \ will be enhanced across several dimensions. First, we plan to broaden its native support for additional DNS implementations, such as MaraDNS and Dnsmasq. Second, we will introduce built-in support for encrypted transport protocols, like DNS over HTTPS (DoH). While DART’s flexible architecture already allows for manual customization of these features, future iterations will focus on abstracting these critical capabilities into high-level primitives, further streamlining the deployment of complex experimental environments.

\bibliographystyle{ACM-Reference-Format}

\begin{thebibliography}{43}


\ifx \showCODEN    \undefined \def \showCODEN     #1{\unskip}     \fi
\ifx \showDOI      \undefined \def \showDOI       #1{#1}\fi
\ifx \showISBNx    \undefined \def \showISBNx     #1{\unskip}     \fi
\ifx \showISBNxiii \undefined \def \showISBNxiii  #1{\unskip}     \fi
\ifx \showISSN     \undefined \def \showISSN      #1{\unskip}     \fi
\ifx \showLCCN     \undefined \def \showLCCN      #1{\unskip}     \fi
\ifx \shownote     \undefined \def \shownote      #1{#1}          \fi
\ifx \showarticletitle \undefined \def \showarticletitle #1{#1}   \fi
\ifx \showURL      \undefined \def \showURL       {\relax}        \fi
\providecommand\bibfield[2]{#2}
\providecommand\bibinfo[2]{#2}
\providecommand\natexlab[1]{#1}
\providecommand\showeprint[2][]{arXiv:#2}

\bibitem[Afek et~al\mbox{.}(2020)]%
        {NXNS2020}
\bibfield{author}{\bibinfo{person}{Yehuda Afek}, \bibinfo{person}{Anat Bremler{-}Barr}, {and} \bibinfo{person}{Lior Shafir}.} \bibinfo{year}{2020}\natexlab{}.
\newblock \showarticletitle{NXNSAttack: Recursive {DNS} Inefficiencies and Vulnerabilities}. In \bibinfo{booktitle}{\emph{29th {USENIX} Security Symposium, {USENIX} Security 2020, August 12-14, 2020}}, \bibfield{editor}{\bibinfo{person}{Srdjan Capkun} {and} \bibinfo{person}{Franziska Roesner}} (Eds.). \bibinfo{publisher}{{USENIX} Association}, \bibinfo{pages}{631--648}.
\newblock


\bibitem[Afek et~al\mbox{.}(2023)]%
        {NRDelegation2023}
\bibfield{author}{\bibinfo{person}{Yehuda Afek}, \bibinfo{person}{Anat Bremler{-}Barr}, {and} \bibinfo{person}{Shani Stajnrod}.} \bibinfo{year}{2023}\natexlab{}.
\newblock \showarticletitle{NRDelegationAttack: Complexity DDoS attack on {DNS} Recursive Resolvers}. In \bibinfo{booktitle}{\emph{32nd {USENIX} Security Symposium, {USENIX} Security 2023, Anaheim, CA, USA, August 9-11, 2023}}, \bibfield{editor}{\bibinfo{person}{Joseph~A. Calandrino} {and} \bibinfo{person}{Carmela Troncoso}} (Eds.). \bibinfo{publisher}{{USENIX} Association}, \bibinfo{pages}{3187--3204}.
\newblock


\bibitem[Biondi(2026)]%
        {scapy}
\bibfield{author}{\bibinfo{person}{Philippe Biondi}.} \bibinfo{year}{2026}\natexlab{}.
\newblock \bibinfo{title}{{Scapy: Interactive packet manipulation program}}.
\newblock
\newblock
\urldef\tempurl%
\url{https://scapy.net/}
\showURL{%
\tempurl}


\bibitem[CloudFront(2026)]%
        {Amazon-CloudFront}
\bibfield{author}{\bibinfo{person}{Amazon CloudFront}.} \bibinfo{year}{2026}\natexlab{}.
\newblock \bibinfo{title}{{Low-Latency Content Delivery Network (CDN) - Amazon CloudFront - Amazon Web Services}}.
\newblock \bibinfo{howpublished}{\url{https://aws.amazon.com/cloudfront/}}.
\newblock


\bibitem[{CZ.NIC}(2023)]%
        {knotresolver}
\bibfield{author}{\bibinfo{person}{{CZ.NIC}}.} \bibinfo{year}{2023}\natexlab{}.
\newblock \bibinfo{title}{{Knot Resolver: A caching full DNS resolver}}.
\newblock
\newblock
\urldef\tempurl%
\url{https://www.knot-resolver.cz/}
\showURL{%
\tempurl}


\bibitem[{Dan Kaminsky}(2008)]%
        {Kaminsky2010}
\bibfield{author}{\bibinfo{person}{{Dan Kaminsky}}.} \bibinfo{year}{2008}\natexlab{}.
\newblock \bibinfo{title}{It’s the End of the Cache as We Know It}.
\newblock \bibinfo{howpublished}{\url{https://www.blackhat.com/presentations/bh-jp-08/bh-jp-08-Kaminsky/BlackHatJapan-08-Kaminsky-DNS08-BlackOps.pdf}}.
\newblock


\bibitem[de~C{\'{a}}mara et~al\mbox{.}(2024)]%
        {CamaraFAUZ24}
\bibfield{author}{\bibinfo{person}{Xabier~S{\'{a}}ez de C{\'{a}}mara}, \bibinfo{person}{Jose~Luis Flores}, \bibinfo{person}{Crist{\'{o}}bal Arellano}, \bibinfo{person}{Aitor Urbieta}, {and} \bibinfo{person}{Urko Zurutuza}.} \bibinfo{year}{2024}\natexlab{}.
\newblock \showarticletitle{Gotham Testbed: {A} Reproducible IoT Testbed for Security Experiments and Dataset Generation}.
\newblock \bibinfo{journal}{\emph{{IEEE} Trans. Dependable Secur. Comput.}} \bibinfo{volume}{21}, \bibinfo{number}{1} (\bibinfo{year}{2024}), \bibinfo{pages}{186--203}.
\newblock


\bibitem[{DNS-OARC}(2026)]%
        {dnsperf}
\bibfield{author}{\bibinfo{person}{{DNS-OARC}}.} \bibinfo{year}{2026}\natexlab{}.
\newblock \bibinfo{title}{{dnsperf: DNS Performance Testing Tool}}.
\newblock
\newblock
\urldef\tempurl%
\url{https://www.dns-oarc.net/tools/dnsperf}
\showURL{%
\tempurl}


\bibitem[Du et~al\mbox{.}(2022)]%
        {SEED2022}
\bibfield{author}{\bibinfo{person}{Wenliang Du}, \bibinfo{person}{Honghao Zeng}, {and} \bibinfo{person}{Kyungrok Won}.} \bibinfo{year}{2022}\natexlab{}.
\newblock \showarticletitle{{SEED} emulator: an internet emulator for research and education}. In \bibinfo{booktitle}{\emph{Proceedings of the 21st {ACM} Workshop on Hot Topics in Networks, HotNets 2022, Austin, Texas, November 14-15, 2022}}. \bibinfo{publisher}{{ACM}}, \bibinfo{pages}{101--107}.
\newblock


\bibitem[Duan et~al\mbox{.}(2024)]%
        {CAMP2024}
\bibfield{author}{\bibinfo{person}{Huayi Duan}, \bibinfo{person}{Marco Bearzi}, \bibinfo{person}{Jodok Vieli}, \bibinfo{person}{David~A. Basin}, \bibinfo{person}{Adrian Perrig}, \bibinfo{person}{Si Liu}, {and} \bibinfo{person}{Bernhard Tellenbach}.} \bibinfo{year}{2024}\natexlab{}.
\newblock \showarticletitle{{CAMP:} Compositional Amplification Attacks against {DNS}}. In \bibinfo{booktitle}{\emph{33rd {USENIX} Security Symposium, {USENIX} Security 2024, Philadelphia, PA, USA, August 14-16, 2024}}. \bibinfo{publisher}{{USENIX} Association}.
\newblock


\bibitem[{EVE-NG Ltd.}(2026)]%
        {EVE-NG}
\bibfield{author}{\bibinfo{person}{{EVE-NG Ltd.}}} \bibinfo{year}{2026}\natexlab{}.
\newblock \bibinfo{title}{{The Emulated Virtual Environment Next Generation ({EVE-NG})}}.
\newblock \bibinfo{howpublished}{\url{https://www.eve-ng.net/}}.
\newblock


\bibitem[Fry et~al\mbox{.}(2023)]%
        {hickory}
\bibfield{author}{\bibinfo{person}{Benjamin Fry} {et~al\mbox{.}}} \bibinfo{year}{2023}\natexlab{}.
\newblock \bibinfo{title}{{Hickory DNS (formerly Trust-DNS): A Rust based DNS client, server, and resolver}}.
\newblock
\newblock
\urldef\tempurl%
\url{https://github.com/hickory-dns/hickory-dns}
\showURL{%
\tempurl}


\bibitem[{GNS3 Technologies}(2026)]%
        {gns3}
\bibfield{author}{\bibinfo{person}{{GNS3 Technologies}}.} \bibinfo{year}{2026}\natexlab{}.
\newblock \bibinfo{title}{Graphical Network Simulator-3 ({GNS3})}.
\newblock \bibinfo{howpublished}{\url{https://www.gns3.com/}}.
\newblock


\bibitem[Heftrig et~al\mbox{.}(2024)]%
        {KeyTrap2024}
\bibfield{author}{\bibinfo{person}{Elias Heftrig}, \bibinfo{person}{Haya Schulmann}, \bibinfo{person}{Niklas Vogel}, {and} \bibinfo{person}{Michael Waidner}.} \bibinfo{year}{2024}\natexlab{}.
\newblock \showarticletitle{The Harder You Try, The Harder You Fail: The KeyTrap Denial-of-Service Algorithmic Complexity Attacks on {DNSSEC}}. In \bibinfo{booktitle}{\emph{Proceedings of the 2024 on {ACM} {SIGSAC} Conference on Computer and Communications Security, {CCS} 2024, Salt Lake City, UT, USA, October 14-18, 2024}}, \bibfield{editor}{\bibinfo{person}{Bo~Luo}, \bibinfo{person}{Xiaojing Liao}, \bibinfo{person}{Jun Xu}, \bibinfo{person}{Engin Kirda}, {and} \bibinfo{person}{David Lie}} (Eds.). \bibinfo{publisher}{{ACM}}, \bibinfo{pages}{497--510}.
\newblock


\bibitem[Heftrig et~al\mbox{.}(2023)]%
        {DgradDNSSEC2023}
\bibfield{author}{\bibinfo{person}{Elias Heftrig}, \bibinfo{person}{Haya Schulmann}, {and} \bibinfo{person}{Michael Waidner}.} \bibinfo{year}{2023}\natexlab{}.
\newblock \showarticletitle{Downgrading {DNSSEC:} How to Exploit Crypto Agility for Hijacking Signed Zones}. In \bibinfo{booktitle}{\emph{32nd {USENIX} Security Symposium, {USENIX} Security 2023, Anaheim, CA, USA, August 9-11, 2023}}. \bibinfo{publisher}{{USENIX} Association}, \bibinfo{pages}{7429--7444}.
\newblock


\bibitem[{Internet Systems Consortium (ISC)}(2026a)]%
        {dig}
\bibfield{author}{\bibinfo{person}{{Internet Systems Consortium (ISC)}}.} \bibinfo{year}{2026}\natexlab{a}.
\newblock \bibinfo{title}{{BIND 9: dig (domain information groper)}}.
\newblock
\newblock
\urldef\tempurl%
\url{https://www.isc.org/bind/}
\showURL{%
\tempurl}


\bibitem[{Internet Systems Consortium (ISC)}(2026b)]%
        {bind}
\bibfield{author}{\bibinfo{person}{{Internet Systems Consortium (ISC)}}.} \bibinfo{year}{2026}\natexlab{b}.
\newblock \bibinfo{title}{{BIND 9: The most widely used DNS software}}.
\newblock
\newblock
\urldef\tempurl%
\url{https://www.isc.org/bind/}
\showURL{%
\tempurl}


\bibitem[Izhikevich et~al\mbox{.}(2022)]%
        {ZDNS2022}
\bibfield{author}{\bibinfo{person}{Liz Izhikevich}, \bibinfo{person}{Gautam Akiwate}, \bibinfo{person}{Briana Berger}, \bibinfo{person}{Spencer Drakontaidis}, \bibinfo{person}{Anna Ascheman}, \bibinfo{person}{Paul Pearce}, \bibinfo{person}{David Adrian}, {and} \bibinfo{person}{Zakir Durumeric}.} \bibinfo{year}{2022}\natexlab{}.
\newblock \showarticletitle{{ZDNS:} a fast {DNS} toolkit for internet measurement}. In \bibinfo{booktitle}{\emph{Proceedings of the 22nd {ACM} Internet Measurement Conference, {IMC} 2022, Nice, France, October 25-27, 2022}}. \bibinfo{publisher}{{ACM}}, \bibinfo{pages}{33--43}.
\newblock


\bibitem[Jeitner and Schulmann(2021)]%
        {Inject2021}
\bibfield{author}{\bibinfo{person}{Philipp Jeitner} {and} \bibinfo{person}{Haya Schulmann}.} \bibinfo{year}{2021}\natexlab{}.
\newblock \showarticletitle{Injection Attacks Reloaded: Tunnelling Malicious Payloads over {DNS}}. In \bibinfo{booktitle}{\emph{30th {USENIX} Security Symposium, {USENIX} Security 2021, August 11-13, 2021}}, \bibfield{editor}{\bibinfo{person}{Michael~D. Bailey} {and} \bibinfo{person}{Rachel Greenstadt}} (Eds.). \bibinfo{publisher}{{USENIX} Association}, \bibinfo{pages}{3165--3182}.
\newblock


\bibitem[Kenneally and Dittrich(2012)]%
        {kenneally2012menlo}
\bibfield{author}{\bibinfo{person}{Erin Kenneally} {and} \bibinfo{person}{David Dittrich}.} \bibinfo{year}{2012}\natexlab{}.
\newblock \showarticletitle{The menlo report: Ethical principles guiding information and communication technology research}.
\newblock \bibinfo{journal}{\emph{Available at SSRN 2445102}} (\bibinfo{year}{2012}).
\newblock


\bibitem[Lantz et~al\mbox{.}(2010)]%
        {Mininet2010}
\bibfield{author}{\bibinfo{person}{Bob Lantz}, \bibinfo{person}{Brandon Heller}, {and} \bibinfo{person}{Nick McKeown}.} \bibinfo{year}{2010}\natexlab{}.
\newblock \showarticletitle{A network in a laptop: rapid prototyping for software-defined networks}. In \bibinfo{booktitle}{\emph{Proceedings of the 9th {ACM} Workshop on Hot Topics in Networks. HotNets 2010, Monterey, CA, {USA} - October 20 - 21, 2010}}. \bibinfo{publisher}{{ACM}}, \bibinfo{pages}{19}.
\newblock


\bibitem[Li et~al\mbox{.}(2023a)]%
        {PhoenixDomain2023}
\bibfield{author}{\bibinfo{person}{Xiang Li}, \bibinfo{person}{Baojun Liu}, \bibinfo{person}{Xuesong Bai}, \bibinfo{person}{Mingming Zhang}, \bibinfo{person}{Qifan Zhang}, \bibinfo{person}{Zhou Li}, \bibinfo{person}{Haixin Duan}, {and} \bibinfo{person}{Qi Li}.} \bibinfo{year}{2023}\natexlab{a}.
\newblock \showarticletitle{Ghost Domain Reloaded: Vulnerable Links in Domain Name Delegation and Revocation}. In \bibinfo{booktitle}{\emph{30th Annual Network and Distributed System Security Symposium, {NDSS} 2023, San Diego, California, USA, February 27 - March 3, 2023}}. \bibinfo{publisher}{The Internet Society}.
\newblock


\bibitem[Li et~al\mbox{.}(2023b)]%
        {Maginot2023}
\bibfield{author}{\bibinfo{person}{Xiang Li}, \bibinfo{person}{Chaoyi Lu}, \bibinfo{person}{Baojun Liu}, \bibinfo{person}{Qifan Zhang}, \bibinfo{person}{Zhou Li}, \bibinfo{person}{Haixin Duan}, {and} \bibinfo{person}{Qi Li}.} \bibinfo{year}{2023}\natexlab{b}.
\newblock \showarticletitle{The Maginot Line: Attacking the Boundary of {DNS} Caching Protection}. In \bibinfo{booktitle}{\emph{32nd {USENIX} Security Symposium, {USENIX} Security 2023, Anaheim, CA, USA, August 9-11, 2023}}, \bibfield{editor}{\bibinfo{person}{Joseph~A. Calandrino} {and} \bibinfo{person}{Carmela Troncoso}} (Eds.). \bibinfo{publisher}{{USENIX} Association}, \bibinfo{pages}{3153--3170}.
\newblock


\bibitem[Li et~al\mbox{.}(2024a)]%
        {DNSBomb2024}
\bibfield{author}{\bibinfo{person}{Xiang Li}, \bibinfo{person}{Dashuai Wu}, \bibinfo{person}{Haixin Duan}, {and} \bibinfo{person}{Qi Li}.} \bibinfo{year}{2024}\natexlab{a}.
\newblock \showarticletitle{DNSBomb: {A} New Practical-and-Powerful Pulsing DoS Attack Exploiting {DNS} Queries-and-Responses}. In \bibinfo{booktitle}{\emph{{IEEE} Symposium on Security and Privacy, {SP} 2024, San Francisco, CA, USA, May 19-23, 2024}}. \bibinfo{publisher}{{IEEE}}, \bibinfo{pages}{4478--4496}.
\newblock


\bibitem[Li et~al\mbox{.}(2024b)]%
        {TuDoor2024}
\bibfield{author}{\bibinfo{person}{Xiang Li}, \bibinfo{person}{Wei Xu}, \bibinfo{person}{Baojun Liu}, \bibinfo{person}{Mingming Zhang}, \bibinfo{person}{Zhou Li}, \bibinfo{person}{Jia Zhang}, \bibinfo{person}{Deliang Chang}, \bibinfo{person}{Xiaofeng Zheng}, \bibinfo{person}{Chuhan Wang}, \bibinfo{person}{Jianjun Chen}, \bibinfo{person}{Haixin Duan}, {and} \bibinfo{person}{Qi Li}.} \bibinfo{year}{2024}\natexlab{b}.
\newblock \showarticletitle{TuDoor Attack: Systematically Exploring and Exploiting Logic Vulnerabilities in {DNS} Response Pre-processing with Malformed Packets}. In \bibinfo{booktitle}{\emph{{IEEE} Symposium on Security and Privacy, {SP} 2024, San Francisco, CA, USA, May 19-23, 2024}}. \bibinfo{publisher}{{IEEE}}, \bibinfo{pages}{4459--4477}.
\newblock


\bibitem[Li et~al\mbox{.}(2025)]%
        {Rebirthday2025}
\bibfield{author}{\bibinfo{person}{Xiang Li}, \bibinfo{person}{Mingming Zhang}, \bibinfo{person}{Zuyao Xu}, \bibinfo{person}{Fasheng Miao}, \bibinfo{person}{Yuqi Qiu}, \bibinfo{person}{Baojun Liu}, \bibinfo{person}{Jia Zhang}, \bibinfo{person}{Xiaofeng Zheng}, \bibinfo{person}{Haixin Duan}, \bibinfo{person}{Zheli Liu}, \bibinfo{person}{Yunhai Zhang}, {and} \bibinfo{person}{Dunqiu Fan}.} \bibinfo{year}{2025}\natexlab{}.
\newblock \showarticletitle{RebirthDay Attack: Reviving {DNS} Cache Poisoning with the Birthday Paradox}. In \bibinfo{booktitle}{\emph{Proceedings of the 2025 {ACM} {SIGSAC} Conference on Computer and Communications Security, {CCS} 2025, Taipei, Taiwan, October 13-17, 2025}}. \bibinfo{pages}{1619--1633}.
\newblock


\bibitem[Liu et~al\mbox{.}(2026a)]%
        {DNSPUN}
\bibfield{author}{\bibinfo{person}{Shiming Liu}, \bibinfo{person}{Yunyi Zhang}, \bibinfo{person}{Ruixuan Li}, \bibinfo{person}{Shiyao Guo}, \bibinfo{person}{Baojun Liu}, \bibinfo{person}{Donghong Sun}, \bibinfo{person}{Yong Ma}, {and} \bibinfo{person}{Linjian Song}.} \bibinfo{year}{2026}\natexlab{a}.
\newblock \showarticletitle{The Trade-off Between Performance and Security: Exploring Vulnerabilities in DNS Task Queue Scheduling}. In \bibinfo{booktitle}{\emph{Proceedings of the 2026 {ACM} {SIGSAC} Conference on Computer and Communications Security, {CCS} 2026, Hague, Netherlands, November 15-19, 2026}}.
\newblock


\bibitem[Liu et~al\mbox{.}(2026b)]%
        {BADDNS}
\bibfield{author}{\bibinfo{person}{Shiming Liu}, \bibinfo{person}{Yunyi Zhang}, \bibinfo{person}{Chaoyi Lu}, \bibinfo{person}{Baojun Liu}, \bibinfo{person}{Shuhan Zhang}, \bibinfo{person}{Donghong Sun}, \bibinfo{person}{Yong Ma}, {and} \bibinfo{person}{Linjian Song}.} \bibinfo{year}{2026}\natexlab{b}.
\newblock \showarticletitle{The Trade-off Between Performance and Security: Exploring Vulnerabilities in DNS Task Queue Scheduling}. In \bibinfo{booktitle}{\emph{Proceedings of the 2026 {ACM} {SIGSAC} Conference on Computer and Communications Security, {CCS} 2026, Hague, Netherlands, November 15-19, 2026}}.
\newblock


\bibitem[Moura et~al\mbox{.}(2021)]%
        {TsuName2021}
\bibfield{author}{\bibinfo{person}{Giovane C.~M. Moura}, \bibinfo{person}{Sebastian Castro}, \bibinfo{person}{John~S. Heidemann}, {and} \bibinfo{person}{Wes Hardaker}.} \bibinfo{year}{2021}\natexlab{}.
\newblock \showarticletitle{TsuNAME: exploiting misconfiguration and vulnerability to DDoS {DNS}}. In \bibinfo{booktitle}{\emph{{IMC} '21: {ACM} Internet Measurement Conference, Virtual Event, USA, November 2-4, 2021}}, \bibfield{editor}{\bibinfo{person}{Dave Levin}, \bibinfo{person}{Alan Mislove}, \bibinfo{person}{Johanna Amann}, {and} \bibinfo{person}{Matthew Luckie}} (Eds.). \bibinfo{publisher}{{ACM}}, \bibinfo{pages}{398--418}.
\newblock


\bibitem[Nagle(2026)]%
        {packetsender}
\bibfield{author}{\bibinfo{person}{Dan Nagle}.} \bibinfo{year}{2026}\natexlab{}.
\newblock \bibinfo{title}{{Packet Sender: Network Utility for TCP, UDP, SSL}}.
\newblock
\newblock
\urldef\tempurl%
\url{https://packetsender.com/}
\showURL{%
\tempurl}


\bibitem[{NLnet Labs}(2026)]%
        {unbound}
\bibfield{author}{\bibinfo{person}{{NLnet Labs}}.} \bibinfo{year}{2026}\natexlab{}.
\newblock \bibinfo{title}{{Unbound: A validating, recursive, and caching DNS resolver}}.
\newblock
\newblock
\urldef\tempurl%
\url{https://www.nlnetlabs.nl/projects/unbound/}
\showURL{%
\tempurl}


\bibitem[{Nokia SRL}(2026)]%
        {containerlab}
\bibfield{author}{\bibinfo{person}{{Nokia SRL}}.} \bibinfo{year}{2026}\natexlab{}.
\newblock \bibinfo{title}{Containerlab: Docker-based Meshed Lab Topologies}.
\newblock \bibinfo{howpublished}{\url{https://containerlab.dev/}}.
\newblock


\bibitem[{NS1}(2026)]%
        {flamethrower}
\bibfield{author}{\bibinfo{person}{{NS1}}.} \bibinfo{year}{2026}\natexlab{}.
\newblock \bibinfo{title}{{FlameThrower: A DNS performance and functional testing utility}}.
\newblock
\newblock
\urldef\tempurl%
\url{https://github.com/NS1/flamethrower}
\showURL{%
\tempurl}


\bibitem[Pan et~al\mbox{.}(2024)]%
        {Loopy2024}
\bibfield{author}{\bibinfo{person}{Yepeng Pan}, \bibinfo{person}{Anna Ascheman}, {and} \bibinfo{person}{Christian Rossow}.} \bibinfo{year}{2024}\natexlab{}.
\newblock \showarticletitle{Loopy Hell(ow): Infinite Traffic Loops at the Application Layer}. In \bibinfo{booktitle}{\emph{33rd {USENIX} Security Symposium, {USENIX} Security 2024, Philadelphia, PA, USA, August 14-16, 2024}}. \bibinfo{publisher}{{USENIX} Association}.
\newblock


\bibitem[Partridge and Allman(2016)]%
        {partridge2016ethical}
\bibfield{author}{\bibinfo{person}{Craig Partridge} {and} \bibinfo{person}{Mark Allman}.} \bibinfo{year}{2016}\natexlab{}.
\newblock \showarticletitle{Ethical considerations in network measurement papers}.
\newblock \bibinfo{journal}{\emph{Commun. ACM}} (\bibinfo{year}{2016}).
\newblock


\bibitem[Pochat et~al\mbox{.}(2019)]%
        {Tranco2023}
\bibfield{author}{\bibinfo{person}{Victor~Le Pochat}, \bibinfo{person}{Tom van Goethem}, \bibinfo{person}{Samaneh Tajalizadehkhoob}, \bibinfo{person}{Maciej Korczynski}, {and} \bibinfo{person}{Wouter Joosen}.} \bibinfo{year}{2019}\natexlab{}.
\newblock \showarticletitle{Tranco: {A} Research-Oriented Top Sites Ranking Hardened Against Manipulation}. In \bibinfo{booktitle}{\emph{26th Annual Network and Distributed System Security Symposium, San Diego, California, USA, February 24-27, 2019}}. \bibinfo{publisher}{The Internet Society}.
\newblock


\bibitem[{PowerDNS}(2026)]%
        {powerdns}
\bibfield{author}{\bibinfo{person}{{PowerDNS}}.} \bibinfo{year}{2026}\natexlab{}.
\newblock \bibinfo{title}{{PowerDNS Recursor}}.
\newblock
\newblock
\urldef\tempurl%
\url{https://www.powerdns.com/}
\showURL{%
\tempurl}


\bibitem[Sacramento(2002)]%
        {Birthday}
\bibfield{author}{\bibinfo{person}{Vagner Sacramento}.} \bibinfo{year}{2002}\natexlab{}.
\newblock \bibinfo{title}{Vulnerability in Requests Control of BIND Versions 4 and 8 Allows DNS Spoofing}.
\newblock
\newblock
\urldef\tempurl%
\url{https://lists.isc.org/pipermail/bind-users/2002November/043141.html}
\showURL{%
\tempurl}


\bibitem[Sanfilippo(2026)]%
        {hping3}
\bibfield{author}{\bibinfo{person}{Salvatore Sanfilippo}.} \bibinfo{year}{2026}\natexlab{}.
\newblock \bibinfo{title}{{hping3: A command-line oriented TCP/IP packet assembler/analyzer}}.
\newblock
\newblock
\urldef\tempurl%
\url{http://www.hping.org/}
\showURL{%
\tempurl}


\bibitem[Wu et~al\mbox{.}(2026)]%
        {Cuckoo}
\bibfield{author}{\bibinfo{person}{Yuxiao Wu}, \bibinfo{person}{Yunyi Zhang}, \bibinfo{person}{Baojun Liu}, {and} \bibinfo{person}{Chaoyi Lu}.} \bibinfo{year}{2026}\natexlab{}.
\newblock \showarticletitle{Should I Trust You? Rethinking the Principle of Zone-Based Isolation DNS Bailiwick Checking}. In \bibinfo{booktitle}{\emph{33th Annual Network and Distributed System Security Symposium, San Diego, California, USA, February 23 - 27, 2026}}. \bibinfo{publisher}{The Internet Society}.
\newblock


\bibitem[Xu et~al\mbox{.}(2023)]%
        {TsuKing2023}
\bibfield{author}{\bibinfo{person}{Wei Xu}, \bibinfo{person}{Xiang Li}, \bibinfo{person}{Chaoyi Lu}, \bibinfo{person}{Baojun Liu}, \bibinfo{person}{Haixin Duan}, \bibinfo{person}{Jia Zhang}, \bibinfo{person}{Jianjun Chen}, {and} \bibinfo{person}{Tao Wan}.} \bibinfo{year}{2023}\natexlab{}.
\newblock \showarticletitle{TsuKing: Coordinating {DNS} Resolvers and Queries into Potent DoS Amplifiers}. In \bibinfo{booktitle}{\emph{Proceedings of the 2023 {ACM} {SIGSAC} Conference on Computer and Communications Security, {CCS} 2023, Copenhagen, Denmark, November 26-30, 2023}}. \bibinfo{publisher}{{ACM}}, \bibinfo{pages}{311--325}.
\newblock


\bibitem[Zhang et~al\mbox{.}(2024)]%
        {ResolverFuzz2024}
\bibfield{author}{\bibinfo{person}{Qifan Zhang}, \bibinfo{person}{Xuesong Bai}, \bibinfo{person}{Xiang Li}, \bibinfo{person}{Haixin Duan}, \bibinfo{person}{Qi Li}, {and} \bibinfo{person}{Zhou Li}.} \bibinfo{year}{2024}\natexlab{}.
\newblock \showarticletitle{ResolverFuzz: Automated Discovery of {DNS} Resolver Vulnerabilities with Query-Response Fuzzing}. In \bibinfo{booktitle}{\emph{33rd {USENIX} Security Symposium, {USENIX} Security 2024, Philadelphia, PA, USA, August 14-16, 2024}}. \bibinfo{publisher}{{USENIX} Association}.
\newblock


\bibitem[Zheng et~al\mbox{.}(2020)]%
        {ForwardFrag2020}
\bibfield{author}{\bibinfo{person}{Xiaofeng Zheng}, \bibinfo{person}{Chaoyi Lu}, \bibinfo{person}{Jian Peng}, \bibinfo{person}{Qiushi Yang}, \bibinfo{person}{Dongjie Zhou}, \bibinfo{person}{Baojun Liu}, \bibinfo{person}{Keyu Man}, \bibinfo{person}{Shuang Hao}, \bibinfo{person}{Haixin Duan}, {and} \bibinfo{person}{Zhiyun Qian}.} \bibinfo{year}{2020}\natexlab{}.
\newblock \showarticletitle{Poison Over Troubled Forwarders: {A} Cache Poisoning Attack Targeting {DNS} Forwarding Devices}. In \bibinfo{booktitle}{\emph{29th {USENIX} Security Symposium, {USENIX} Security 2020, August 12-14, 2020}}, \bibfield{editor}{\bibinfo{person}{Srdjan Capkun} {and} \bibinfo{person}{Franziska Roesner}} (Eds.). \bibinfo{publisher}{{USENIX} Association}, \bibinfo{pages}{577--593}.
\newblock


\end{thebibliography}

\appendix

\section{Open Science and Ethics}

\noindent \textbf{Open science.} To foster reproducibility in DNS protocol analysis, we have open-sourced \tool \ and its comprehensive suite of pre-configured testing environments. This contribution aims to provide a standardized benchmark for future research in DNS security and compliance.

\noindent \textbf{Ethics.} Our experiments strictly adhere to the principles outlined in the Menlo Report~\cite{kenneally2012menlo} and established best practices for internet measurement~\cite{partridge2016ethical}. To mitigate potential risks, all replicated experiments were conducted within a strictly contained environment, ensuring no attack traffic escaped to the public Internet. Furthermore, to prevent misuse, we have open-sourced only the reproducible testing frameworks and environment configurations, intentionally excluding the specific attack scripts.


\section{DNS Builder workflow}

\begin{figure*}[h]
    \centering
    \includegraphics[width=0.9\linewidth, trim=0cm 0.3cm 0cm 0.2cm, clip]{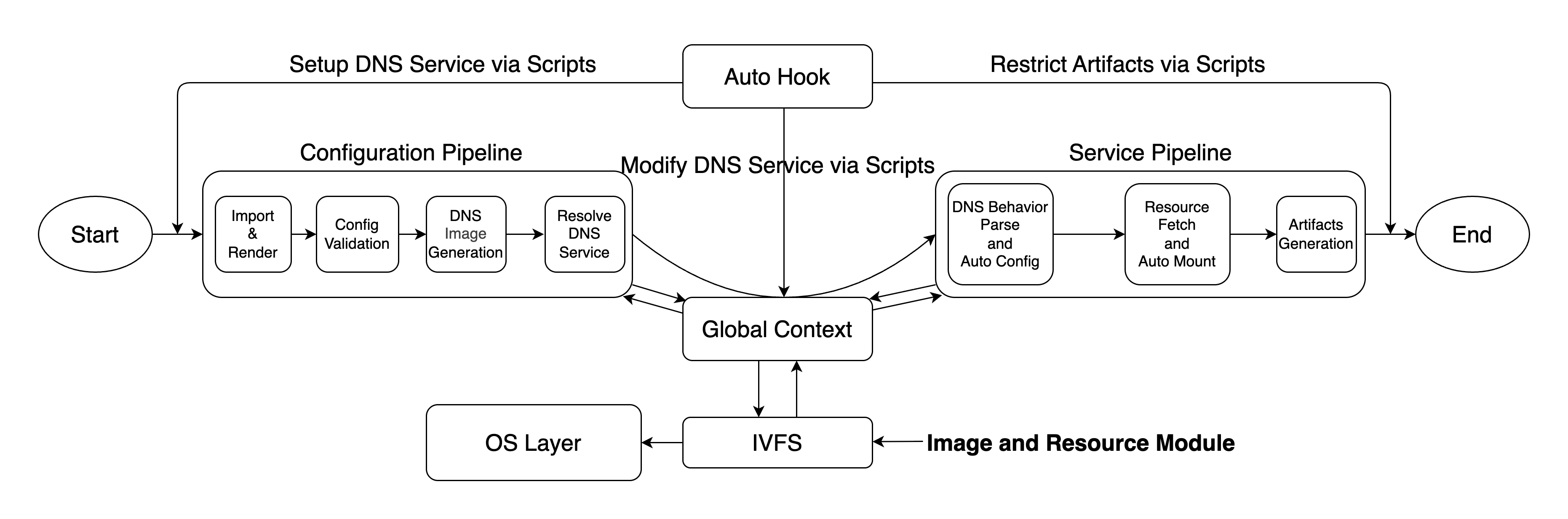}
    \caption{DNS Builder workflow.}
    \label{fig:dnsb_workflow}
\end{figure*}

The DNS Builder environment construction process comprises two primary pipelines: the Configuration Pipeline and the Service Pipeline, as shown in Figure~\ref{fig:dnsb_workflow}.

The Configuration Pipeline transforms user-defined YAML specifications into an Intermediary Representation (IR) stored within a Global Context. During this stage, DNS Builder pre-processes the primary configuration, handling imports, rendering templates, and performing schema validation. It subsequently stages image assets into a temporary internal virtual file system (IVFS) and commits pointers to the global context. Finally, it resolves service-specific configurations, including policy merging, topology planning, and variable substitution.

The Service Pipeline consumes the global context to generate deployable Docker compose artifacts. It begins by rendering service-specific behavior logic into the temporary IVFS, followed by fetching all resource dependencies and automatically generating Docker volume mappings. The staged content is then committed and synchronized to the user-specified output directory.

Throughout this process, \tool \ provides an Auto Hook mechanism, allowing researchers to inject custom scripts into specific lifecycle phases. This enables advanced programmatic customization, such as the automated generation of large-scale DNS nodes and the dynamic synthesis of complex service configurations.

    
\section{Automated DNSSEC signing and deployment workflow}

\begin{figure*}[h]
    \centering
    \includegraphics[width=0.9\linewidth, trim=0cm 0.3cm 0cm 0.2cm, clip]{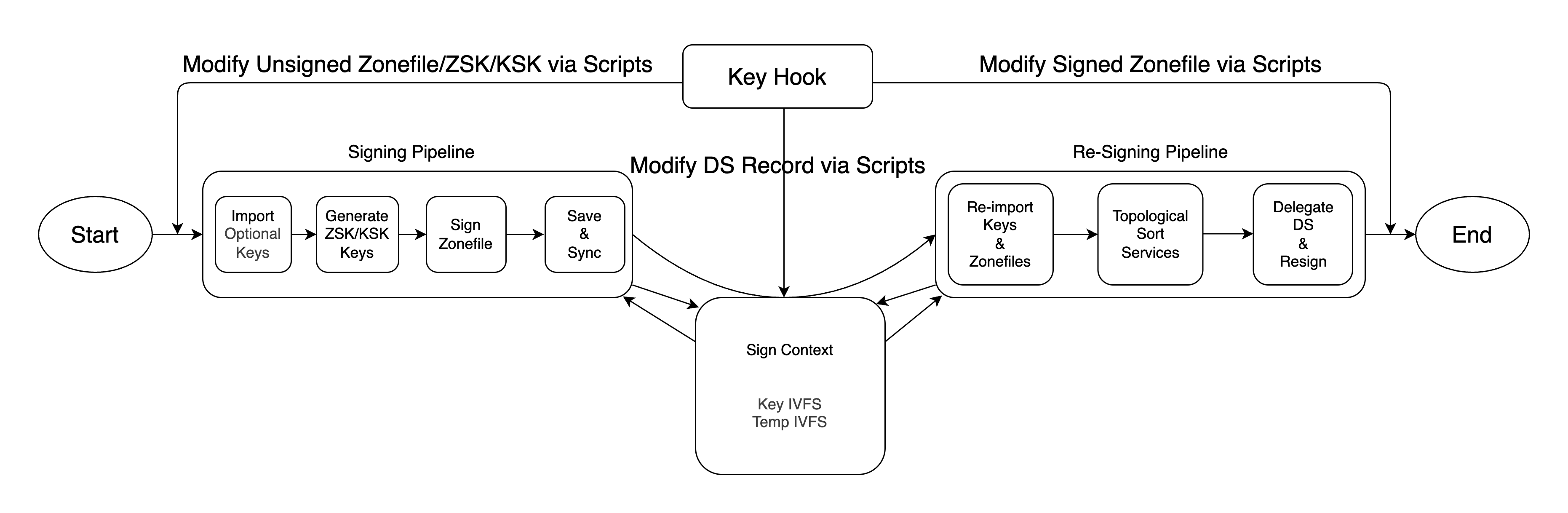}
    \caption{Automated DNSSEC signing and deployment workflow.}
    \label{fig:dnssec_workflow}
\end{figure*}

DNS Builder simplifies the deployment of DNSSEC-enabled environments by automating the generation and referencing of DNSKEYs, performing zone-file signing, and establishing a complete chain of trust originating from the root. This significantly reduces the operational overhead and complexity typically associated with manual DNSSEC configuration. The workflow is driven by two specialized pipelines: the Signing Pipeline and the Re-signing Pipeline, as shown in Figure~\ref{fig:dnssec_workflow}.

The Signing Pipeline is responsible for the parallel initial signing of zone files across multiple DNS services. Specifically, DNS Builder initializes an isolated environment to stage the zone files for each service. It then automatically generates or imports existing DNSKEYs to perform the signing process within this secure abstraction. Finally, the resulting DNSKEYs and Delegation Signer (DS) records are committed to the Key IVFS, while the signed zone files are synchronized to the temporary IVFS for subsequent deployment.

The Re-signing Pipeline is tasked with constructing the end-to-end DNSSEC chain of trust across the isolated environment. During this phase, DNS Builder intercepts the signing workflow to identify the zones managed by each service and performs a topological sort based on their hierarchical relationships. This ensures that Delegation Signer (DS) records are propagated correctly from child to parent zones. Finally, it triggers a re-signing process for any services whose zone files were modified during the propagation.

Throughout the signing process, \tool \ provides a Key Hook mechanism, enabling researchers to inject custom cryptographic operations. This allows for the programmatic introduction of security anomalies, such as forged DNSKEYs, invalid DS records, or tampered RRSIGs.

\section{Configuration example of TsuKing}


\begin{verbatim}
# ------------------------------------- #
# Tsuking Attack Experiment Configuration
# ${vars.DRS_NUM} DNS Resolver System
# every system has 1 ingress, ${vars.DRS_WIDTH} egress
# 
# ------------------------------------ #

# --- Base Configuration ---
name: tsuking
inet: 10.66.0.0/24 # The IPv4 subnet

vars:
  # --- Experiment Parameters ---
  # Number of DNS Resolver System generated all
  DRS_NUM: 16
  # Number of egress Resolver per DNS Resolver System
  DRS_WIDTH: 1

  URL: git://to-what-you-place-you-server-py

  # Default path for the Unbound configuration file
  UCFG: /usr/local/etc/unbound/unbound.conf

images:
  # --- Base Images ---
  unbound:
    ref: unbound:1.17.1
  python39:
    ref: python:3.9.0

builds:
  # --- Base Node Definitions & Templates ---
  auth:
    # Custom Auth DNS
    image: python39
    volumes:
      - ${vars.URL}#server.py:/usr/local/etc/server.py
    command: "python3 /usr/local/etc/server.py"
    
  victim:
    image: python39
    command: "tail -f /dev/null"
    
  fl:
    # Template for Ingress Resolvers
    image: unbound
    volumes:
      - ${vars.URL}#weak.conf:${vars.UCFG}
    build: false # Marks this as a blueprint/template
    
  rl:
    # Template for Egress Resolvers
    # stub example.com to custom auth
    image: unbound
    volumes:
      - ${vars.URL}#weak.conf:${vars.UCFG}
    behavior: example.com stub auth
    build: false # Marks this as a blueprint/template

auto:
  # --- Dynamic Topology Generation Script ---
  # Auto generate all the DRS
  setup: |
    drs = config.get('vars', {}).get('DRS_NUM', 16)
    width = config.get('vars', {}).get('DRS_WIDTH', 1)
    
    # Loop to create DRS_NUM DRS
    for i in range(drs):
      config['builds'][f'fl-{i}'] = {
        'ref': 'fl', # Inherits from the 'fl'
        'behavior': f'. forward {",".join(
            [
            f"rl-{i}-{j}" for j in range(width)
            ]
        )}'
      }
      
    # Loop to create 'width' number or egress
    for i in range(drs):
      for j in range(width):
        config['builds'][f'rl-{i}-{j}'] = {
          'ref': 'rl' # Inherits from the 'rl'
        }
\end{verbatim}

\section{Test cases description}
\label{ap:test_case}

Following established methodologies, we designed three categories of test cases: Functional, Edge-case, and DNSSEC-related. 

The functional test cases are designed to evaluate the core capabilities of resolvers, comprising the following four scenarios:

\begin{itemize}
    \item ns\_delegation: This case verifies the resolver’s ability to correctly navigate multi-level redirections and authoritative NS delegations, ensuring it can traverse deep hierarchical chains to resolve the final answer.
    \item response\_no\_qr: This case confirms that the resolver strictly validates the QR bit of incoming messages, ensuring that packets expected from authoritative servers are correctly identified as responses (QR=1) rather than queries.
    \item dname\_query: This case evaluates the resolver’s compliance with RFC 6672 by verifying its ability to correctly process DNAME records.
    \item rd\_flag\_clear: This case verifies that when receiving a query without the Recursion Desired (RD) flag, the resolver correctly enforces access policy by returning a REFUSED response accompanied by Extended DNS Error (EDE) code 20 (Not Recursive).

\end{itemize}

Edge-case test cases evaluate resolver behaviors in specific scenarios mandated by RFCs. We designed three such scenarios:
\begin{itemize}
    \item invalid\_domain: This case verifies whether the resolver correctly identifies the ``.invalid'' TLD as a reserved special-use name.
    \item test\_domain: This case ensures the resolver correctly identifies the ``.test'' TLD as a reserved name under RFC 6761. The resolver should immediately return an error (e.g., NXDOMAIN) and refrain from issuing redundant upstream queries to the root or authoritative servers.
    \item onion\_domain: This case validates compliance with RFC 7686 by ensuring the resolver identifies ``.onion'' as a special-use TLD.
\end{itemize}

DNSSEC-related test cases evaluate the resolver’s adherence to specifications. We designed three test cases:
\begin{itemize}
    \item cd\_flag\_query: This case validates that when a query is received with the Checking Disabled (CD) bit set, the resolver bypasses DNSSEC validation and returns all relevant records (including potentially invalid ones) rather than filtering them with a SERVFAIL response.
    \item edns\_key\_tag: This case verifies that a non-validating resolver correctly handles the ``edns-key-tag'' option (RFC 8145). According to EDNS specifications, if the resolver does not perform DNSSEC validation, it must transparently copy and forward this option in its response rather than dropping it.
    \item multi\_tsig\_records: This case validates that the resolver strictly enforces the single-signature rule of TSIG (RFC 2845). Upon receiving a message containing multiple TSIG records, the resolver must identify the protocol violation and immediately return a FORMERR (Format Error) response.
\end{itemize}




\end{document}